\documentstyle[proceedings,numreferences,epsfig]{balkanschool}

\begin{opening}
\title{NON-EQUILIBRIUM AND COLLECTIVE FLOW EFFECTS IN RELATIVISTIC HEAVY ION COLLISIONS}
\author{T. GAITANOS, H.H. WOLTER}
\institute{Sektion Physik der Universit\"at M\"unchen \\ 
Am Coulombwall 1, D-85748 Garching, Germany}
\author{C. FUCHS}
\institute{Institut f\"ur Theoretische Physik der Universit\"at T\"ubingen \\ 
Auf der Morgenstelle 14, T\"ubingen, Germany}
\end{opening}
\runningtitle{NON-EQUILIBRIUM EFFECTS}
\begin{document}
\newcommand{\lvecp}{\raisebox{1.77ex}{\mbox{\tiny$\leftarrow$}}\hspace{-0.7em} \partial}
\newcommand{\p}{\partial}
\newcommand{\ls}{\left(}
\newcommand{\rs}{\right)}
\newcommand{\beq}{\begin{equation}}
\newcommand{\eeq}{\end{equation}}
\newcommand{\beqa}{\begin{eqnarray}}
\newcommand{\eeqa}{\end{eqnarray}}
\newcommand{\Gs}{\Gamma_s}
\newcommand{\Go}{\Gamma_0}
\newcommand{\dGo}{\frac{ \partial{\Gamma_{0}} }{ \partial{\rho_0} }}
\newcommand{\dGs}{\frac{ \partial{\Gamma_{s}} }{ \partial{\rho_0} }}
\newcommand{\ddGo}{\frac{ \partial^2 {\Gamma_{0}} }{ \partial{\rho_0^2} }}
\newcommand{\ddGs}{\frac{ \partial^2 {\Gamma_{s}} }{ \partial{\rho_0^2} }}
\newcommand{\jj}{(j^{\lambda} j_{\lambda})}
\newcommand{\dj}{(\partial^{\mu}j^{\alpha})j_{\alpha}}
\newcommand{\dnj}{(\partial^{\nu}j^{\alpha})j_{\alpha}}
\newcommand{\dr}{(\partial^{\mu} \rho_s)}
\newcommand{\dnr}{(\partial^{\nu} \rho_s)}
\newcommand{\su}{\sum_{j=1}^{A \cdot N}}
\newcommand{\ex}{e^{ \frac{R_j^2}{\sigma^2}}}
\newcommand{\uj}{(u_j^{\alpha} j_{\alpha})}
\section{Introduction}
Heavy ion physics opens the unique opportunity to explore the nuclear equation of 
state (EOS) far away from saturation, i.e. at high densities and at non zero 
temperatures. Investigations of nuclear matter under such extreme conditions 
are crucial for the understanding of the universe, its evolution and the formation 
of the elements, and the evolution of massive stars, supernovae and neutron stars. 

The intensive study of heavy ion collisions has became possible by the introduction 
of new generations of accelerators and detectors, which allow to study collisions of 
heavy nuclei from Fermi energies up to high relativistic energies of a few hundred 
$GeV$ per nucleon with an almost complete characterization 
of the reaction products of the space-time events. 

Inspite of the theoretical efforts over the last three decades (see ref. \cite{review}) 
the determination of the EOS in heavy ion reactions is still an object of 
current debate. Comparisons with experiments seem to favor a nuclear EOS with an 
incompressibility of $K \approx 230~MeV$ and with a moderate density dependence 
at high densities (soft EOS). In particular, a difficultly in the determination of the 
nuclear EOS is the fact that the nuclear matter is to a large extent far away 
from local equilibrium in a heavy ion collision. These non-equilibrium effects 
are a main feature of energetic nuclear collisions, and govern the dynamics of the 
reaction with relaxation times comparable to the compression phase of the 
process \cite{gait1}. These effects originate from an anisotropy 
in momentum space and lead to effective fields which are different from 
the ground state \cite{gait2}. 

In this contribution we discuss the origin of non-equilibrium features, 
their influence on the nuclear matter EOS and the consequences when 
drawing conclusions on the nuclear EOS from comparisons with experiments 
in terms of collective flow effects.

\section{The nuclear EOS: from nuclear matter to heavy ion collisions}
The theoretical concept of the description of nuclear matter (NM) is given in an effective 
relativistic quantum field 
theory, Quantumhadrodynamics (QHD) \cite{qhd}. It is formulated in terms of baryon 
and meson fields as the important non-perturbative degrees of freedom of QCD. A 
relativistic treatment is attractive since it yields a natural description of characteristic 
features of NM, such as the appereance of strong attractive scalar and repulsive 
vector fields which naturally account for the saturation mechanism and for the strong 
spin-orbit potential. Relativity also leads to effectively energy dependent fields, which 
are crucial for nucleon-nucleus reactions and also heavy ion collisions.
The effective Lagrangian of QHD involving baryon and meson fields is given by (for 
brevity, only scalar $\sigma$ and vector $\omega^{\mu}$ meson fields are considered here) 
\begin{eqnarray}
{\cal L}_{QHD} & = & {\cal L}_{B} + {\cal L}_{M} + {\cal L}_{int}
\nonumber\\
{\cal L}_{B} & = & \overline{\Psi} ( i\gamma_{\mu} \partial^{\mu} - M ) \Psi 
\nonumber\\
{\cal L}_{M} & = & \frac{1}{2} \left( 
                                \partial_{\mu} \sigma \partial^{\mu} \sigma - 
                                m_{\sigma}^{2} \sigma^{2} 
                                \right) 
                  - \frac{1}{4} F_{\mu\nu} F^{\mu\nu} 
                   - m_{\omega}^{2} \omega_{\alpha} \omega^{\alpha}
\nonumber\\
{\cal L}_{int} & = & \Gamma_{\sigma} \overline{\Psi} \Psi \sigma - 
                     \Gamma_{\omega} \overline{\Psi} \gamma_{\alpha} \Psi \omega^{\alpha}
\label{L_QHD}
\qquad ,
\end{eqnarray}
with $F^{\mu\nu}=\partial^{\mu} \omega^{\nu} - \partial^{\nu} \omega^{\mu}$ 
the field tensor. 
The mesons $\sigma$ and vector $\omega^{\mu}$ are coupled in a minimal way to the 
nucleons via effective couplings $\Gamma_{\sigma}$ and $\Gamma_{\omega}$, 
respectively. 

Different approximations have been used for the solution of eq. (\ref{L_QHD}) in 
nuclear matter. The most popular one is the Hartree or Relativistic Mean Field (RMF) 
approach where the meson fields are treated classically \cite{qhd}. A more realistic 
treatment is the Dirac-Brueckner-Hartree-Fock (DBHF) theory, in which exchange terms and 
higher order correlations  within a ${\cal T}$-Matrix or ladder approximation are 
taken into account \cite{dbhf}. The success of DBHF theory was in the unified description 
of nucleon-nucleon (NN) scattering and saturation properties of nuclear matter which 
was not possible in non-relativistic approaches, except by 
including $3$-body forces. The nucleonic mean 
field potential is given in terms of self energies by 
\begin{equation}
\Sigma^{DBHF}(p,p_{F}) = \Sigma_{s}(p,p_{F}) - \gamma_{\mu}\Sigma^{\mu}(p,p_{F})
\label{Sigma_DB}
\qquad ,
\end{equation}
with scalar and vector components $\Sigma_{s}, \Sigma^{\mu}$ depending on density 
$\rho(p_{F})$ and for energies higher that the Fermi energy on momentum $p$. 
For NM the DBHF theory is parameter free, 
since it is based on a model for the bare NN interaction given by boson exchange 
potentials \cite{mach}. This is in contrast to 
phenomenological approaches, where several parameters are adjusted to NM 
saturation properties. 

The application of DBHF theory to finite nuclei has been formulated in a Density Dependent 
Hadronic (DDH) field theory \cite{lenske}, where the self energies are parametrized in 
Hartree form by 
\begin{equation}
\Sigma_{s} = \Gamma_{s}(p,p_{F}) \rho_{s}(p_{F})
\quad\mbox{,} \quad 
\Sigma_{\mu} = \Gamma_{0}(p,p_{F})j_{\mu}
\label{Sigma_DDH}
\qquad ,
\end{equation}
with density and momentum dependent vertex functions $\Gamma_{s,0}$ 
that replace the constant values of the RMF approach 
($\Gamma_{s,0} \rightarrow \Gamma^{2}_{\sigma,\omega}/m^{2}_{\sigma,\omega}$). They 
effectively contain exchange and correlation effects of the DBHF theory. 
\begin{figure}[t]
\begin{center}
\unitlength1cm
\begin{picture}(9,5.)
\put(0,-0.5){\makebox{\epsfig{file=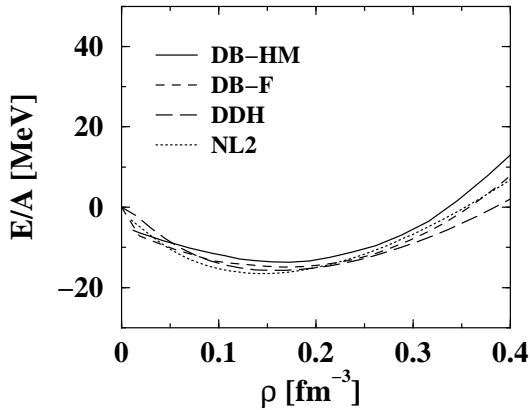,width=7.0cm}}}
\end{picture}
\caption{\label{Fig1} {\it Nuclear matter EOS for different models: (solid) DBHF from 
\protect\cite{dbhm}, (dashed) DBHF from \protect\cite{dbf}, (long-dashed) RMF with 
density dependent couplings \protect\cite{ddh} and (dotted) Walecka model 
(NL2 parametrization) \protect\cite{nl2} }}
\end{center}
\end{figure}
Fig.~\ref{Fig1} compares the density dependence of the ground state EOS obtained from 
different models, which show a similar density behavior around saturation, but signifigant 
differences at high densities. The density dependence of the EOS 
at supra-normal densities can be regarded as an extrapolation which is 
tested in heavy ion collisions, where high compressions of the matter are reached. 

However, 
the determination of the nuclear matter EOS from heavy ion collisions is only indirect, 
because it contains fields corresponding to non-equilibrium colliding matter. 
Thus one does not, in fact, 
{\it see} the equilibrated nuclear matter EOS directly in heavy ion collisions.

\begin{figure}[t]
\begin{center}
\unitlength1cm
\begin{picture}(9,5.3)
\put(0,-0.5){\makebox{\epsfig{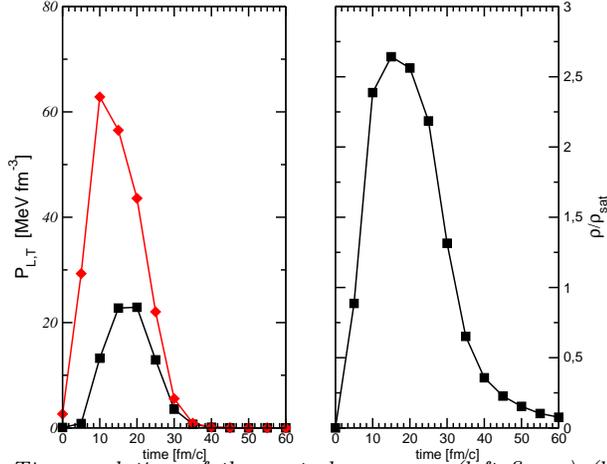}}}
\end{picture}
\caption{\label{Fig2} {\it Time evolution of the central pressures (left figure) 
(longitudinal $P_{L}$ and transversal $P_{T}$) and the central density (right figure) 
for a central $Au+Au$ heavy ion collision at $E_{lab}=0.4~A.GeV$ }}
\end{center}
\end{figure}

Equilibrated 
nuclear matter is characterized by an isotropic spherical momentum distribution. This 
is not the case in the dynamical situations of heavy ion collisions. This is seen in 
Fig.~\ref{Fig2} in terms of the transversal and longitudinal components of the 
{\it local} central pressure, obtained in calculations with a transport equation, 
as described in the next section. 
The anisotropy in the pressure components indicates 
that the matter is {\it locally} not equilibrated, except at the late stage of the 
procces ($t > 30~fm/c$). Indeed, the relaxation 
times are large and comparable with the compression phase of the collision (see 
time evolution of the central density in \ref{Fig2}). Therefore, it is important 
to first study possible influences of phase space anisotropies on the nuclear 
matter EOS before making comparisons with experiments.

The investigation of non-equilibrium effects on the level of the effective mean fields 
has been done by considering two models: a Local Density and a 
Colliding Nuclear Matter 
approximations, denoted as LDA and CNM respectively. In the LDA the EOS and the 
corresponding fields are those of equilibrated matter decribed by an 
isotropic momentum 
distribution. The mean field 
$\Sigma(p_{F},p)=\Sigma_{s}(p_{F},p)-\gamma_{\mu}\Sigma^{\mu}(p_{F},p)$
depends on a density $\rho(E_{F})$ and for energies 
above the Fermi energy $E_{F}$ on momentum $p$ relative to the rest system of
nuclear matter. The density 
dependence has been taken from phenomenological (Walecka, DDH) or microscopic (DBHF) 
models. The momentum dependence is also naturally provided from the 
DBHF theory, whereas in the DDH approach it has been adjusted to the energy dependence 
of the Schr\"odinger equivalent optical potential in nucleon-nucleus scattering 
\cite{typel2}. The fields of the non-linear Walecka model (NL2) do not depend on energy.

\begin{figure}[t]
\begin{center}
\unitlength1cm
\begin{picture}(9,2.5)
\put(0,-0.5){\makebox{\epsfig{file=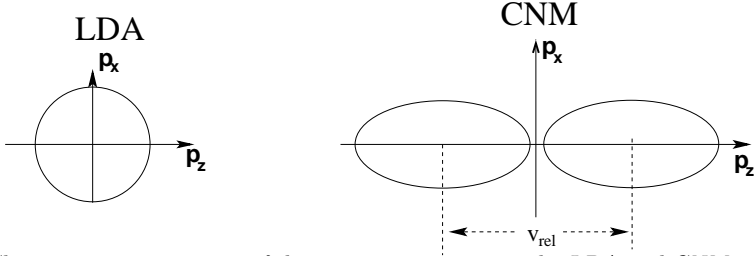,width=10.0cm}}}
\end{picture}
\caption{\label{Fig3} {\it Shematic representation of the momentum space in the 
LDA and CNM models.}}
\end{center}
\end{figure}

In the CNM model the anisotropy of momentum space is modeled by that of two 
covariant Fermi-ellipsoids, i.e. two interpenetrating currents of nuclear matter 
(see Fig.~\ref{Fig3}) 
$f_{CNM}(p_{F_{1}},p_{F_{2}},v_{rel})=f_1(p_{F_{1}})+f_2(p_{F_{2}})+\delta f(p_{F_{1,2}},v_{rel})$ 
where the last term takes Pauli effects into account. This representation 
of phase space anisotropies is motivated from microscopic calculations of 
heavy ion collisions \cite{essler}. 
The mean field is then given (e.g. for the scalar part) approximately by
\begin{equation}
\Sigma^{CNM}(p,p_{F};v_{rel}) = \Sigma_{s1}(\Lambda^{-1}_{1}p,p_{F_{1}}) + 
\Sigma_{s2}(\Lambda^{-1}_{2}p,p_{F_{2}}) + 
\delta \Sigma(\Lambda^{-1}_{1,2}p,p_{F_{1,2}},v_{rel})
\label{Sigma_CNM}
\qquad ,
\end{equation}
i.e. as a superposition of self energies of the nucleon with respect 
to the two currents taking into 
account the Lorentz transformations $\Lambda_{1,2}$ for the momenta and the Pauli 
correction in the overlap region ($\delta \Sigma$). This ensures 
the correct limit to the LDA of one Fermi-sphere, i.e. 
$CNM \stackrel{v_{rel} \rightarrow 0}{\longrightarrow} LDA$. 
As a further approximation an  
average of the $p$-dependent quantities $\Sigma_{s,0,v,\cdots}^{CNM}$ over the CNM 
configuration leads to fields depending only on the configuration parameters, i.e. 
the two Fermi momenta $p_{F_{1,2}}$ and the relative velocity $v_{rel}$ \cite{sehn}. 

The energy-momentum tensor $T^{\mu\nu}$ is given in terms of the CNM 
fields by \cite{gait2,sehn}
\begin{eqnarray}
T^{\mu\nu} & = & <p^{*\mu}p^{*\nu}/E^{*}>_{CNM}-<p^{*\mu}\Sigma^{CNM\nu}/E^{*}>_{CNM} 
\nonumber\\
& - &\frac{1}{2}g^{\mu\nu} \left[ <\Sigma_{s}m^{*}/E^{*}>_{CNM}
-<p^{*\lambda}\Sigma^{CNM}_{\lambda}/E^{*}>_{CNM}  \right]
\label{T_CNM}
\qquad ,
\end{eqnarray}
where $< \cdots >_{CNM}$ means the average over the CNM momentum space. The energy 
per nucleon $E^{CNM}(\rho_{tot},v_{rel})=T^{00}/\rho_{tot}-M$ depends on the total 
invariant density $\rho_{tot}=\sqrt{j_{tot\alpha}j_{tot}^{\alpha}}$ and the relative 
velocity $v_{rel}$. However, the total energy $E^{CNM}$ contains contributions from 
the energy of the relative motion of the two currents. A meaningful discussion of 
non-equilibrium effects with respect to the ground state EOS should be based on the 
binding energy and thus it is reasonable to subtract 
contributions from the relative motion of the two currents 
$E_{rel}$. $E_{rel}$ can be relativistically defined as the 
difference of the kinetic energy of the CNM system and the corresponding 
LDA configuration at twice the subsystem density $2 \rho$
\begin{equation}
E_{rel}(\rho_{tot},v_{rel}) = <\sqrt{p^{*2}+m^{*2}}-m^{*}>_{CNM} - 
<\sqrt{p^{*2}+m^{*2}}-m^{*}>_{v_{rel}=0}
\label{E_REL}
\qquad .
\end{equation}
By definition $E_{rel}$ contains the kinetic energy arising from the separation 
of the Fermi-spheres in 
momentum space. It accounts for the interaction between the two currents 
by the presence of the effective mass in eq. (\ref{E_REL}). 
One should note that eq. (\ref{E_REL}) is the natural definition of a relativistic 
kinetic energy of the relative motion of two {\it interacting} currents and contains, 
in particular, the correct non-relativistic limit \cite{gait2}.

\begin{figure}[t]
\begin{center}
\unitlength1cm
\begin{picture}(9,8)
\put(0,-0.5){\makebox{\epsfig{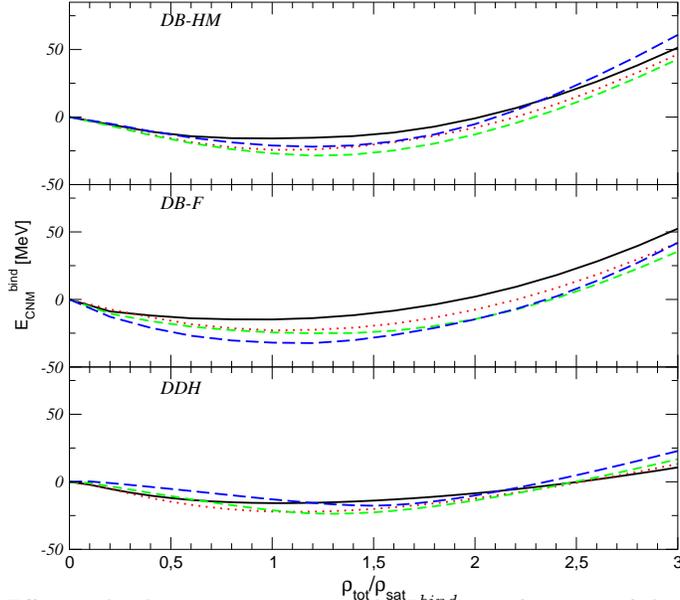}}}
\end{picture}
\caption{\label{Fig4} {\it Effective binding energy per particle $E_{CNM}^{bind}$ 
as function of the total density $\rho_{tot}$ (normalized to the saturation density 
$\rho_{sat}$) at different relative velocities $v_{rel}$ 
(solid: $v_{rel}=0$, dotted: $v_{rel}=0.2$, dashed: $v_{rel}=0.4$ 
and long-dashed: $v_{rel}=0.6$) 
for the DBHF theory 
from Refs. \protect\cite{dbhm} (DB-HM) and \protect\cite{dbf} (DB-F) and for the 
DDH approach \protect\cite{ddh}. }}
\end{center}
\end{figure}
Now one is able to construct an effective bound energy of colliding nuclear 
matter $E_{CNM}^{bind}(\rho_{tot},v_{rel})$ 
\begin{equation}
E_{CNM}^{bind}(\rho_{tot},v_{rel}) = \frac{T^{00}-E_{rel}}{\rho_{tot}}-M
\label{E_BIND}
\qquad ,
\end{equation}
which relates colliding to ground state nuclear matter. $E_{CNM}^{bind}$ is displayed in 
Fig.~\ref{Fig4} for different models. The case $v_{rel}=0$ is the correct limit 
for LDA. With increasing $v_{rel}$ 
we observe a {\it softening of the effective EOS} in terms of an increase of the 
binding energy at $\rho_{tot} \sim \rho_{sat}$ and a decrease of $E^{bind}_{CNM}$ at 
high densities ($\rho_{tot} >> \rho_{sat}$) with respect to the ground state EOS 
(solid curves in Fig.~\ref{Fig4}). The softening of the effective EOS occurs in all 
the models investigated and can be explained by (a) a decrease of the fields 
$\Sigma^{CNM}_{s,0,v,\cdots}$ with density and momentum, see eq. (\ref{Sigma_CNM}), 
(b) a reduction of the Fermi pressure (not shown here) in transverse direction and 
(c) an enlarged scalar attraction with increasing relative velocity \cite{gait2}. 
It is important that the differences between the non-equilibrium and the 
ground state EOS are of the same magnitude as the differences of the  
ground state EOS for different models, see 
e.g. Fig.~\ref{Fig1}. Thus we conclude that one should take non-equilibrium effects into 
account when determining the 
EOS in heavy ion collisions. In the next section we investigate the non-equilibrium 
effects in transport calculations of heavy ion collisions in terms of collective flow 
observables.

\section{Heavy ion collisions}
Heavy ion collisions have been described by covariant transport equations of a 
Boltzmann type (details can be found in refs. \cite{nl2,TPE} ):
\begin{eqnarray}
& &  \left[ \left( m^* \p_{x}^\mu m^* - p^{*\nu} \p_{x}^\mu p^{*}_{\nu} \right) \p^{p}_\mu 
- \left( m^* \p_{p}^\mu m^* - p^{*\nu} \p_{p}^\mu p^{*}_{\nu} \right) \p^{x}_\mu 
\right] f(x,{\bf p})
\nonumber\\
& & =  \frac{1}{2} \int \frac{d^4 p_{2}}{E^{*}_{p_{2}}(2\pi)^3} \frac{d^4 p_{3}}{E^{*}_{p_{3}}(2\pi)^3}
             \frac{d^4 p_{4}}{E^{*}_{p_{4}}(2\pi)^3} W(pp_2|p_3 p_4)   
(2\pi)^4 \delta^4 \left(p + p_{2} -p_{3} - p_{4} \right) 
\nonumber\\  
& & \times \; \Big[ \: f(x,{\bf p}_3) f(x,{\bf p}_4) 
                      \ls 1-  f(x,{\bf p}) \rs \ls 1- f(x,{\bf p}_2) \rs - 
\nonumber \\  
& & \; \; \; \; \; \:
   f(x,{\bf p}) f(x,{\bf p}_2) 
                \ls 1-  f(x,{\bf p}_3) \rs \ls 1- f(x,{\bf p}_4) \rs \: \Big] 
\label{RBUU} 
\quad ,
\end{eqnarray}
with effective quantities $p^{*\mu}=p^{\mu}-\Sigma^{\mu}$ and $m^{*}=M-\Sigma_{s}$. 
It describes the evolution of the phase space density $f(x,p)$ under the influence of 
a mean field in the left hand side of eq. (\ref{RBUU}) and on the right hand side of 
$2$-body collisions 
determined by energy and isospin dependent NN-cross sections 
$\sigma_{NN} \sim W(pp_{2}|p_{3}p_{4})$ for the procces 
$p+p_{2} \longrightarrow p_{3}+p_{4}$. The Pauli principle is taken into account 
for the final states by the terms $(1-f)$. The NN-cross section is taken 
from empirical NN-scattering data. Also inelastic channels are taken into account 
as the production of $\Delta$ and $N^{*}$ resonances with their decay in $1$- and 
$2$-pion channels \cite{gait1,gait3}. The EOS enters into the theoretical 
descriptions of heavy ion reactions via the self energies.

We have applied the approximations discussed in the previous section, the LDA and 
CNM approaches, in transport calculations using eq. (\ref{RBUU}). In the LDA model 
the ground state EOS enters directly into the transport equation neglecting 
non-equilibrium effects, whereas in the CNM model the effective EOS including the 
momentum space anisotropies is considered in eq. (\ref{RBUU}). In the latter case 
the CNM parameters, i.e. the invariants $\rho_{1},~\rho_{2},~v_{rel}$, are determined directly 
from the phase space density $f(x,p)$ at each position and in each time step of the simulation. 

\begin{figure}[t]
\begin{center}
\unitlength1cm
\begin{picture}(9,6.)
\put(0,-0.5){\makebox{\epsfig{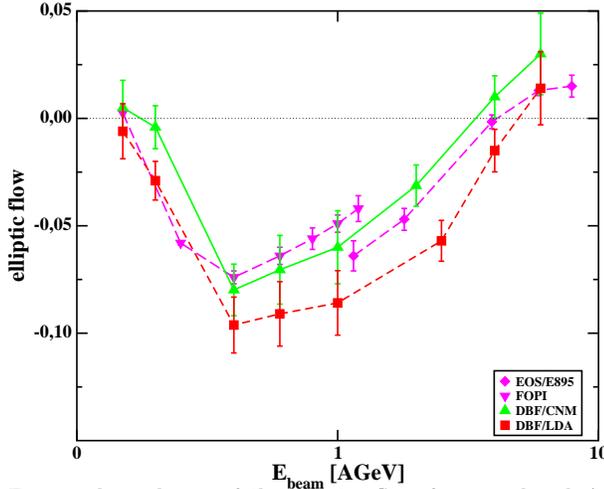}}}
\end{picture}
\caption{\label{Fig5} {\it  Energy dependence of the elliptic flow for 
peripheral $Au+Au$ collisions at energies from the Fermi energy to AGS energies. 
Calculations with the DBHF model \protect\cite{dbf} are shown in the LDA and 
CNM approximations and compared to data \protect\cite{fopi}. }}
\end{center}
\end{figure}
We discuss the results of transport simulations in terms of collective flows, which 
have been found to be sensitive to the density and, particularly, momentum dependence 
of the mean field \cite{daniel00}. The collective flow is characterized by a Fourier 
series of the azimuthal distribution of nucleons for given rapidity $y$ and total 
transverse momentum $p_{t}$, 
$N(\phi,y,p_{t})=N_{0}(1+v_{1}cos\phi+v_{2}cos2\phi \cdots)$, 
where $v_{1}$ and $v_{2}$ are called transverse and elliptic flow, respectively. 
As an example, Fig.~\ref{Fig5} shows the energy dependence of the 
elliptic flow at mid-rapidity ($|\Delta y| \leq 0.15$) 
for peripheral $Au+Au$ collisions. With the 
cut in the rapidity $y=\frac{E+p_{z}}{E-p_{z}}$ one selects particles 
from the hot and compressed matter in the central region of the reaction, from 
which one intends to extract information on the nuclear EOS. 
Thus the elliptic flow 
describes the dynamics of the compressed {\it fireball} perpendicular to the 
beam direction. It probes the high density behavior of the nucleonic 
fields, i.e. their stiffness, which gives rise to a squeeze-out of the compressed matter 
perpendicular 
to the beam direction. An enhancement of the elliptic flow is correlated to a 
stiffer EOS. The calculations in Fig.~\ref{Fig5} were performed with one of the models 
of the previous section, i.e. the DBHF results of the T\"ubingen group \cite{dbf} 
treated in the LDA and in the CNM approximations. 
The two different treatments {\it which are based on 
identical nuclear forces for ground state matter} yield significantly different 
results 
for the elliptic flow. Since this observable is particularly sensitive to the EOS at 
high densities \cite{daniel,daniel00}, these effects are most pronounced here. 
In the LDA approach 
one would have excluded the underlying EOS from the comparison to data as too stiff. The 
more consistent treatment of momentum space anisotropies on the level of the effective 
interaction leads to the net softening of the effective EOS discussed above 
and restores the agreement with experimental data. Similar effects 
are observed 
for other models, e.g. those of the previous section, and other components of 
collective flow not shown here \cite{gait1,gait3}. 

\begin{figure}[t]
\begin{center}
\unitlength1cm
\begin{picture}(9,5.)
\put(-2.0,-0.5){\makebox{\epsfig{file=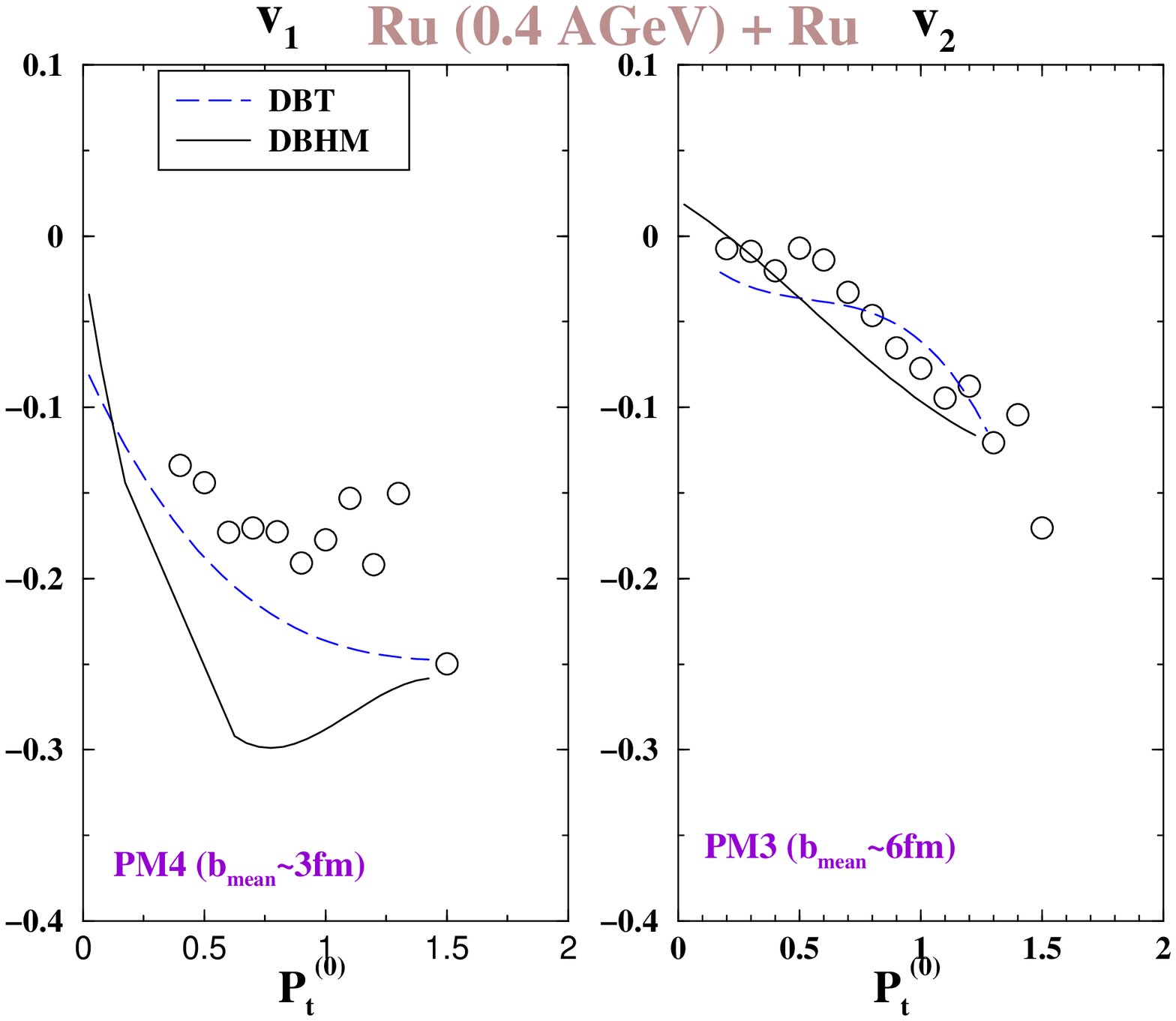,width=6.5cm}}}
\put(5.0,-0.5){\makebox{\epsfig{file=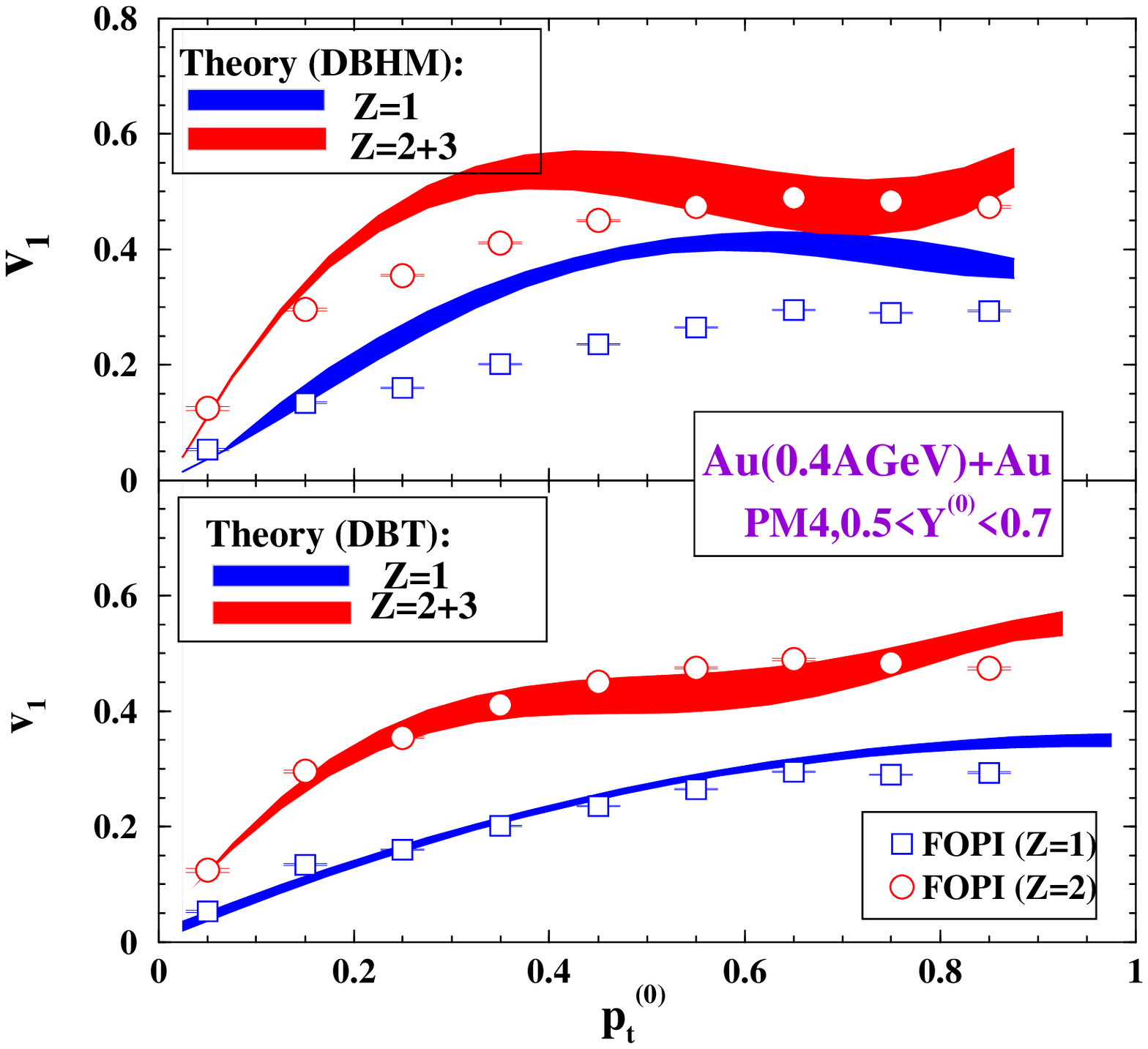,width=6.0cm}}}
\end{picture}
\caption{\label{Fig6} {\it  Transverse momentum 
($p_{t}^{(0)}=\sqrt{p_{x}^{2}+p_{y}^{2}}/p_{t}^{proj}$) dependence of the collective 
flow projected in ($v_{1}$) and out of ($v_{2}$) the reaction plane. (Left) Comparisons 
for $Ru+Ru$ and (right) for $Au+Au$ collisions at $0.4~A.GeV$ beam energy. The figure  
on the right shows the quantity $v_{1}$ for protons ($Z=1$) and fragments ($Z=2$). 
The data (symbols) are from ref. \protect\cite{fopi} and the theoretical calculations 
were performed within the DBHF theory of refs. \protect\cite{dbf,dbhm} in the CNM model. }}
\end{center}
\end{figure}
Other examples of collective flow effects are shown in Fig.~\ref{Fig6} with respect to 
the transverse momentum dependence $p_{t}^{(0)}$. High $p_{t}^{(0)}$-values correspond to 
highly energetic particles originating from the compressed matter 
in an early stage of the reaction and thus directly carry information about the 
high density behavior of the EOS. The DBT calculations with a relatively soft EOS 
also yield a smoother $p_{t}^{(0)}$-dependence of the collective flows 
closer to the data than the DBHM calculations. This effect, which is observed also for other energies, 
is consistent with Figs.~\ref{Fig4} and \ref{Fig5} indicating that a moderate dependence 
of the nuclear matter EOS at high densities seems more realistic (similar effects have been found 
with the phenomenological DDH model).

\section{Conclusions}
The early and high density phases of relativistic heavy ion collisions are to a large 
extent 
governed by highly anisotropic phase space configurations far from local 
equilibrium, which 
can be approximated by counter-streaming or colliding nuclear matter 
configurations. 
In this contribution we discussed the implications for the effective 
EOS which occurs 
in such non-equilibrium configurations and found them to be important. 
The anisotropy in momentum space leads to a 
softening of the 
non-equilibrium EOS with essential deviations compared to the ground state EOS. 
CNM calculations with different effective interactions for the mean field have 
shown the general features of this effect. 
In calculations of energetic heavy ion collisions, where the mean 
field were 
considered in the local density (LDA) and in the more realistic CNM 
approximations, the effective softening of the non-equilibrium EOS was observable 
in terms of collective flow effects. The differences in the flow signals are found 
to be comparable to those of the nuclear matter EOS for different models. We conclude that for 
a reliable determination 
of the nuclear matter equation of state from heavy ion collisions the non-equilibrium 
features of the phase space must be taken into account on the level of the effective 
fields. The results seem to be consistent 
with a moderate dependence on momentum of the nuclear matter EOS at high densities. 
With the successful 
applications of microscopic models to nuclear matter, finite nuclei  and heavy ion 
collisions one expects to 
move towards a unification of the description of very different nuclear systems and to a 
determination of the EOS of nuclear matter.

{\small

}
\end{document}